\documentclass[12pt]{article}

\usepackage{epsfig,graphicx}
\newcommand{\re}{\mbox{Re}}

\newcommand{\be}{\begin{equation}}
\newcommand{\ee}{\end{equation}}
\newcommand{\bea}{\begin{eqnarray}}
\newcommand{\eea}{\end{eqnarray}}
\newcommand{\wtd}{\widetilde}

\begin{document}
\begin{titlepage}
\begin{flushright}
FERMILAB-PUB-04-330-T\\
MCTP-04-63 \\
\end{flushright}
\vspace{0.5cm}
\begin{center}
{\Large \bf Some Phenomenology of Intersecting \\ D-Brane Models} \\
\vspace{1cm} \renewcommand{\thefootnote}{\fnsymbol{footnote}}
{\large Gordon L. Kane$^1$\footnote[1]{Email: gkane@umich.edu},
 Piyush Kumar$^1$\footnote[2]{Email: kpiyush@umich.edu},
Joseph D. Lykken$^2$\footnote[3]{Email: lykken@fnal.gov},
 Ting T. Wang$^1$\footnote[4]{Email: tingwang@umich.edu}} \\
\vspace{1.cm} \renewcommand{\thefootnote}{\arabic{footnote}} {\it
1. Michigan Center for Theoretical Physics \\ Ann
  Arbor, MI 48109, USA \\
2. Fermi National Accelerator Laboratory\\ P.O. Box 500, Batavia, IL 60510, USA}\\
\end{center}
\vspace{0.5cm}


\begin{abstract}
We present some phenomenology of a new class of intersecting D-brane
models. Soft SUSY breaking terms for these models are calculated in the
$u$--moduli dominant SUSY breaking approach (in type
$IIA$). In this case, the dependence of the soft terms on the
Yukawas and Wilson lines drops out. These soft terms have a
different pattern compared to the usual heterotic string models.
Phenomenological implications for dark matter are discussed.
\end{abstract}

\vspace*{1cm}
\end{titlepage}


\section{Introduction}

One of the goals of string phenomenology is to explain/predict
features of low energy physics - both qualitatively and
quantitatively. We are still far from that elusive goal. To make
progress we think it is essential to build more and more
realistic string models and to study their phenomenological
features. Until a few years ago, only the heterotic string was
considered a serious candidate for providing the unified theory of
fundamental interactions. For a sample of heterotic string model
building, see \cite{heterotic}. Developments in the past
few years have shown that type $I$ and type $II$ strings provide
us with new classes of $\mathcal{N}=1$, $D=4$ vacua, with new
avenues for model building. In addition, the concept of D-branes
has provided us with a better understanding of type $I$ (or
equivalently type $IIB$ orientifold) string theory. It has
recently become evident that intersecting D-brane models offer
excellent opportunities for string phenomenology. In fact, these
developments have been collectively dubbed as the second string
(phenomenology) revolution \cite{Ibanez:1999bn}.

This paper is devoted to the detailed study of a particular class
of models based on type $II$ string theory compactifications on
Calabi-Yau manifolds with D\textit{p}-branes wrapping intersecting
cycles on the compact space. This approach to string model
building is distinguished by its computability and simplicity,
together with very appealing phenomenological possibilities. In
these models, gauge interactions are confined to D-branes. Chiral
fermions are open strings which are stretched between two
intersecting branes. They are localized at the brane
intersections. If certain conditions are satisfied, there will be
massless scalars associated with the chiral fermions such that we
have $\mathcal{N}=1$ supersymmetry in the effective field theory.
Because of these attractive features, intersecting brane model
building has drawn considerable attention in recent years and
several semi-realistic models with an SM or MSSM like spectrum
have been constructed \cite{Ibanez:2001jhep, realistic}.

To test these approximate clues and to begin to test string
theory, only reproducing the SM particle content is not enough.
Numerical predictions must be made. In addition, a successful
theory should not just explain existing data, it must also make
predictions which can be tested in future experiments. For the
brane models, if supersymmetry exists and is softly broken, soft
SUSY breaking terms can calculated and tested by future
experimental measurements. A fair amount of work on the low-energy
effective action of intersecting D-brane models has been done. The
stability of these kind of models has been discussed in
\cite{Blumenhagen:2001te}. The question of tree level gauge
couplings, gauge threshold corrections and gauge coupling
unification has been addressed in
\cite{Shiu:1998pa,Cvetic:2002qa,Lust:2003ky,Antoniadis:2000en,
Cremades:2002te,Blumenhagen:2003jy}.
Yukawa couplings and higher point scattering have been studied in
\cite{Cremades:2003qj,Cremades:2004wa,Cvetic:2003ch,Abel:2003yx}.
Some preliminary results for the K\"{a}hler metric have been
obtained in \cite{Kors:2003wf}. A more complete derivation of the
K\"{a}hler metric directly from open/closed string scattering
amplitudes has been done in \cite{Lust:2004cx}, which we use in
this paper. The question of supersymmetry breaking has also been
addressed in such models
\cite{Camara:2003ku,Lust:2004fi,Lust:2004dn,Marchesano:2004th,Cvetic:2004xx},
using techniques of flux compactification in type $IIB$ string
theory. The basic idea is that turning on background closed string
3-form fluxes can provide a source of supersymmetry breaking.
Although an exciting idea, it is not yet complete in all
respects, see \cite{Font:2004cx}.

In this paper, we have taken a more phenomenological approach,
parametrizing the effects of supersymmetry breaking in a more model
independent manner and examining the consequences. Our main goal here is to use the
results of \cite{Lust:2004cx} to calculate and analyze effective
low energy soft supersymmetry breaking terms. We also look at some of their
dark matter applications. Applications to collider phenomenology
will be dealt with in future work. Our main purpose in this paper is to
move the string constructions of this approach closer to broad
phenomenological applications. While we were writing the
manuscript, \cite{Lust:2004dn} appeared, which has a large overlap
in the computation of the soft terms.

The paper is organized as follows. In section \ref{idb:set:gen},
we briefly review the intersecting brane model constructions. Then
in section \ref{idb:set:model}, we describe a brane setup in
detail, for which we will study the soft terms.  This brane setup
was first introduced in \cite{Cremades:2003qj}. We compute the
soft terms under the assumption of section  \ref{idb:set:soft}, in
the $u$-moduli SUSY breaking scenario. We compute the
\emph{general} formulas for the soft terms so that they can be
applied to a large class of type $IIA/B$ brane setups, including
the flux compactification approach
\cite{Camara:2003ku,Lust:2004fi,Lust:2004dn,
Marchesano:2004th,Cvetic:2004xx}. To understand opportunities
better, we apply these general formulas to a particular brane
setup, and study three particular points in the parameter space.
One of these gives a $\wtd W$ LSP, the second gives a $\wtd H$ LSP
and the third gives a mixed $\wtd B$-$\wtd H$ LSP. The three
points represent an almost generic feature of the parameter space
of these intersecting brane models if one requires a light gluino,
which amounts to reducing fine-tuning as suggested in
\cite{Kane:1998im}. In section \ref{idb:set:phen}, we discuss the
phenomenological implications of the above model: the structure of
soft terms, spectrum, gauge unification, issues of flavor and
phase and in particular, the consequences for cosmology. We
conclude in section \ref{idb:sec:conc}. Some technical details are
provided in the Appendix.

\section{General construction of intersecting brane models.}
\label{idb:set:gen}
In this section, we will briefly review the basics of constructing
these models.  More comprehensive treatments can be found in
\cite{Aldazabal:2000cn,Aldazabal:2001jmp,Blumenhagen:2000jhep,
Blumenhagen:2001jhep,Ott:2003yv}.
The setup is as follows - we consider type $IIA$
string theory compactified on a six dimensional manifold
$\mathcal{M}$. It is understood that we are looking at the large
volume limit of compactification, so that perturbation theory is
valid. In general, there are $K$ stacks of intersecting D6-branes
filling four dimensional Minkowski spacetime and wrapping internal
homology 3-cycles of $\mathcal{M}$. Each stack $P$ consists of
$N_P$ coincident D6 branes whose worldvolume is $M_4 \times
{\Pi}_P$, where ${\Pi}_P$ is the corresponding homology class
of each 3-cycle. The closed string degrees of freedom reside in
the entire ten dimensional space, which contain the geometric
scalar moduli fields of the internal space besides the
gravitational fields. The open string degrees of freedom give rise
to the gauge theory on the D6-brane worldvolumes, with gauge
group ${\Pi}_P \,U(N_{P})$. In addition, there are
open string modes which split into states with both ends on the
same stack of branes as well as those connecting different stacks
of branes. The latter are particularly interesting. If for
example, the 3-cycles of two different stacks, say ${\Pi}_P$ and
${\Pi}_Q$ intersect at a single point in $\mathcal{M}$, the
lowest open string mode in the Ramond sector corresponds to a
chiral fermion localized at the four dimensional intersection of
$P$ and $Q$ transforming in the bifundamental of $U(N_{P}) \times
U(N_{Q})$ \cite{Berkooz:1996npb}. The net number of left handed
chiral fermions in the $a b$ sector is given by the intersection
number $I_{P Q} \equiv [{\Pi}_{P}] \cdot [{\Pi}_{Q}]$.

The propagation of massless closed string RR modes on the compact
space $\mathcal{M}$ under which the D-branes are charged, requires
some consistency conditions to be fulfilled. These are known as
the $RR$ tadpole-cancellation conditions, which basically means
that the net $RR$ charge of the configuration has to vanish
\cite{Uranga:2001npb}. In general, there could be additional RR
sources such as orientifold planes or background fluxes. So they
have to be taken into account too. Another desirable constraint
which the models should satisfy is $\mathcal{N} =1$ supersymmetry.
Imposing this
constraint on the closed string sector requires that the internal
manifold $\mathcal{M}$ be a Calabi-Yau manifold. We will see
shortly that imposing the same constraint on the open string
sector leads to a different condition.

A technical remark on the practical formulation of these models is
in order. Till now, we have described the construction in type
$IIA$ string theory. However, it is also possible to rephrase the
construction in terms of type $IIB$ string theory. The two
pictures are related by T-duality. The more intuitive picture of
type $IIA$ intersecting D-branes is converted to a picture with
type $IIB$ D-branes having background magnetic fluxes on their
world volume. It is useful to remember this equivalence as it
turns out that in many situations, it is more convenient to do
calculations in type $IIB$.

Most of the realistic models constructed in the literature involve
toroidal (${T^6}$) compactifications or orbifold/orientifold
quotients of those. In particular, orientifolding introduces O6
planes as well as mirror branes wrapping 3-cycles which are
related to those of the original branes by the orientifold action.
For simplicity, the torus (${T^6}$) is assumed to be factorized
into three 2-tori, i.e ${T^6}$ = $T^2 \times T^2 \times T^2$. Many
examples of the above type are known, especially with those
involving orbifold groups - i) $Z_2 \, \times \,Z_2$
\cite{Cvetic:2001tj} ii) $Z_4 \, \times \,
Z_2$ \cite{Honecker:2003npb}, iii) $Z_4$
\cite{Blumenhagen:2003jhep}, iv) $Z_6$ \cite{Honecker:2004th},
etc.

\section{A local MSSM-like model}\label{idb:set:model}
In order to make contact with realistic low energy physics while
keeping supersymmetry intact, we are led to consider models which
give rise to the chiral spectrum of the MSSM. It has been shown in
\cite{Ibanez:2001jhep} that this requires us to perform an
orientifold twist. A stack of $N_P$ D6 branes wrapping a 3-cycle
not invariant under the orientifold projection will yield a
$U(N_P)$ gauge group, otherwise we get a real $(SO(2N_P))$ or
pseudoreal $(USp\,(2N_P))$ gauge group.

Using the above fact, the brane content for an MSSM-like chiral
spectrum with the correct intersection numbers has been presented
in \cite{Cremades:2003qj}. Constructions with more than four stacks of
branes can be found in \cite{Kokorelis:2003jr}. In the simplest case,
there are four stacks of branes which
give rise to the initial gauge group : $U(3)_a \times Sp(2)_b
\times U(1)_c \times U(1)_d$, where $a,b,c\,\&\,d$ label the
different stacks. The intersection numbers $I_{PQ} = [{\Pi}_P]
\cdot [{\Pi}_Q]$ between a D6-brane stack $P$ and a D6-brane stack
$Q$ is given in terms of the 3-cycles $[{\Pi}_P]$ and $[{\Pi}_Q]$,
which are assumed to be factorizable.
\begin{equation}
[{\Pi}_P] \equiv
[(n_P^1,m_P^1)\otimes(n_P^2,m_P^2)\otimes(n_P^3,m_P^3)]
\end{equation}
where $(n_{P}^i,m_{P}^i)$ denote the wrapping numbers on the
$i^{th}$ 2-torus.The $O6$ planes are wrapped on 3-cycles :
\begin{equation}
[{\Pi}_{O6}] = \bigotimes_{r=1}^3[(1,0)]^r
\end{equation}
\begin{table}
\begin{center}
\begin{tabular}[c]{|c|c|c|c|c|c|}
\hline Stack  & Number of Branes & Gauge Group &
$(n_{\alpha}^1,m_{\alpha}^1)$ & $(n_{\alpha}^2,m_{\alpha}^2)$ &
$(n_{\alpha}^3,m_{\alpha}^3)$\\
\hline $Baryonic$ & $N_a =3$ & $U(3)= SU(3) \times U(1)_a$ &
$(1,0)$ & $(1/{\rho},3{\rho})$ & $(1/{\rho},-3{\rho})$
\\
$Left$ & $N_b =1$ & $USp(2)\cong SU(2)$ & $(0,1)$ & $(1,0)$ & $(0,-1)$\\
$Right$ & $N_c =1$ & $U(1)_c$ & $(0,1)$ & $(0,-1)$ & $(1,0)$\\
$Leptonic$ & $N_d =1$ & $U(1)_d$ & $(1,0)$ & $(1/{\rho},3{\rho})$
& $(1/{\rho},-3{\rho})$
\\
\hline
\end{tabular}
\end{center}
\caption {\small Brane content for an MSSM-like spectrum. The
mirror branes $a^*,b^*,c^*,d^*$ are not shown. $\rho$ can take
values 1, 1/3. For concreteness, we take $\rho =1$ for calculating
the soft terms. However, the parameter space for the soft terms
remains the same for both $\rho =1$ and
$\rho=1/3$.}\label{idb:tab:wr}
\end{table}
Note that for stack $b$, the mirror brane $b^*$ lies on top of
$b$. So even though $N_b = 1$, it can be thought of as a stack of
two D6 branes, which give an $USp(2) \cong SU(2)$ group under the
orientifold projection.

The brane wrapping numbers are shown in Table \ref{idb:tab:wr} and
the chiral particle spectrum from these intersecting branes are shown in
Table \ref{idb:tab:spec}.
\begin{table}[ht]
\begin{center}
\begin{tabular}{|c|c|c|c|c|c|c||c|}\hline
fields& sector& I& $SU(3)_c\times SU(2)_L$& $U(1)_a$&
  $U(1)_c$&$U(1)_d$&$U(1)_Y$ \\ \hline
$Q_L$& $(a,b)$& 3& $(3,2)$  & 1& 0& 0& 1/6 \\ \hline
$U_R$& $(a,c)$& 3& $(3,1)$  &-1& 1& 0&-2/3 \\ \hline
$D_R$& $(a,c^*)$& 3& $(3,1)$&-1&-1& 0& 1/3 \\ \hline
  $L$& $(d,b)$& 3& $(1,2)$  & 0& 0& 1&-1/2 \\ \hline
$E_R$& $(d,c^*)$& 3& $(1,1)$& 0&-1&-1& 1   \\ \hline
$N_R$& $(d,c)$& 3& $(1,1)$  & 0& 1&-1& 0   \\ \hline
$H_u$& $(b,c)$& 1& $(1,2)$  & 0&-1& 0& 1/2 \\ \hline
$H_d$& $(b,c^*)$& 1& $(1,2)$& 0& 1& 0&-1/2 \\ \hline
\end{tabular}
\end{center}
\caption{The MSSM spectrum from intersecting branes. The hypercharge
normalization is given by $Q_Y=
\frac{1}{6}Q_a-\frac{1}{2}Q_c-\frac{1}{2}Q_d$. }
\label{idb:tab:spec}
\end{table}

\subsection{Getting the MSSM}
The above spectrum is free of chiral anomalies. However, it has an
anomalous $U(1)$ given by $U(1)_a$ + $U(1)_d$. This anomaly is
canceled by a generalized Green-Schwarz mechanism
\cite{Aldazabal:2001jmp}, which gives a Stuckelberg mass to the
$U(1)$ gauge boson. The two nonanomalous $U(1)$s are identified
with $(B-L)$ and the third component of right-handed weak isospin
$U(1)_R$ \cite{Cremades:2003qj}. In orientifold models, it could
sometimes happen that some nonanomalous $U(1)$s also get a mass
by the same mechanism \cite{Ibanez:2001jhep}, the details of which
depend on the specific wrapping numbers. It turns out that in the
above model, two massless $U(1)$s survive. In order to break the two
$U(1)$s down to $U(1)_Y$, some fields carrying non-vanishing lepton
number but neutral under $U(1)_Y$ are assumed to develop vevs. This can also be
thought of as the geometrical process of brane recombination
\cite{Marchesano:2004th,Cremades:2002jhep}.

\subsection{Global embedding and supersymmetry breaking}

As can be checked from Table 1, the brane content by itself does
not satisfy the $RR$ tadpole cancellation conditions :
\begin{equation}
\sum_{\alpha}
([{\Pi}_{\alpha}]+[{\Pi}_{\alpha^*}])=32\,[{\Pi}_{O6}]
\end{equation}
\noindent Therefore, this construction has to be embedded in a
bigger one, with extra $RR$ sources included. There are various
ways to do this such as including hidden D-branes or adding
background closed string fluxes in addition to the open string
ones. As a bonus, this could also give rise to spontaneous
supersymmetry breaking. With extra D-branes, one might consider
the possibility of gaugino condensation in the hidden sector
\cite{Cvetic:2003yd}. Alternatively, one could consider turning on
background closed string $NS$-$NS$ and $RR$ fluxes which generate
a non-trivial effective superpotential for moduli, thereby
stabilizing many of them
\cite{Camara:2003ku,Lust:2004fi,Cvetic:2004xx}.

In this paper, we will leave open the questions of actually
embedding the above model in a global one and the mechanism of
supersymmetry breaking. We shall assume that the embedding has
been done and also only \emph{parametrize} the supersymmetry
breaking, in the spirit of
\cite{Kaplunovsky:1993rd,Brignole:1997dp}. We are encouraged because there
exists a claim of a concrete mechanism for the global embedding of
(the T-dual of) this model as well as supersymmetry breaking
\cite{Marchesano:2004th}.

\subsection{Exotic matter and $\mu$
problem}\label{idb:set:model:exo}

The above local model is very simple in many respects, especially
with regard to gauge groups and chiral matter. However, it also
contains exotic matter content which is non-chiral. These
non-chiral fields are related to the untwisted open string moduli
- the D-brane positions and Wilson lines. The presence of these
non-chiral fields is just another manifestation of the old moduli
problem of supersymmetric string vacua. However, it has been argued
\cite{Lust:2004dn,Gorlich:2004qm} that mass terms for the
above moduli can be generated by turning on a $F$- theory 4-form
flux. One then expects that a proper understanding of this problem will
result in a stabilization of all the moduli. As explained in
\cite{Marchesano:2004th}, there could be $\mathcal{N}=1$
embeddings of this local model in a global construction. This
requires additional D-brane sectors and background closed string
3-form fluxes. The other D-brane sectors add new gauge groups as
well as chiral matter, some of which could be charged under the MSSM
gauge group. This may introduce chiral exotics in the spectrum, an
undesirable situation. However, many of these exotics uncharged under
the MSSM gauge group can be made
to go away by giving vevs to scalars parametrizing different flat
directions. In this paper, we assume that there exists an embedding such that
there are no chiral exotics charged under the MSSM. Such exotics can
cause two types of problems. It is of course essential that no states exist
that would already have been observed. It seems likely that can be
arranged. In addition, states that would change the RGE running and details of
the calculations have to be taken into account eventually.

The higgs sector in the local model arises from strings stretching
between stacks $b$ and $c$. However, the net chirality of the $bc$
sector is zero, since the intersection number $I_{bc}$ is zero.
The higgs sector in the above model has a $\mu$ term, which
has a geometrical interpretation. The real part of the $\mu$
parameter corresponds to the separation between stacks $b$ and $c$
in the first torus, while the imaginary part corresponds to a
Wilson line phase along the 1-cycle wrapped on the first torus.
These correspond to flat directions of the moduli space. Adding
background closed string fluxes may
provide another source of $\mu$ term \cite{Camara:2003ku}, which
will lift the flat direction in general. Thus, the effective $\mu$
term relevant for phenomenology is determined by the above
factors and the problem of obtaining an electroweak scale $\mu$
term from a fundamental model remains open. In this paper,
therefore, we will not attempt to calculate $\mu$, and fix it by
imposing electroweak symmetry breaking (EWSB). It is important to study
further the combined effect of the several contributions to $\mu$ and to
EWSB.

\subsection{Type IIA - type IIB equivalence}
As mentioned earlier, it is useful to think about this model in
terms of its T-dual equivalent. In type $IIB$, we are dealing with
D9 branes wrapped on $T^2 \times T^2 \times T^2$ with an open
string background magnetic flux $\mathcal{F}^j$ turned on. Therefore the
D9-branes
have in general mixed Dirichlet and Neumann boundary
conditions. The flux has two parts - one coming from the
antisymmetric tensor $(b^j)$ and the other from the gauge flux
$(F^j)$ so that :
\begin{equation}
{\mathcal{F}}^j = b^j + 2\pi{\alpha}'\,F^j
\end{equation}
\noindent The above compactification leads to the following closed
string K\"{a}hler and complex structure moduli, each of which are
three in number for this model:
\begin{equation}
T^{'j} = b^j + iR_1^{'j}R_2^{'j}\,\sin({\alpha'}^j); \;\;\; U^{'j}=
\frac{R_2^{'j}}{R_1^{'j}}\, e^{i{\alpha'}^j};\;\;j=1,2,3.
\end{equation}

\noindent where $R_1^{'j}$ and $R_2^{'j}$ are lengths of the
basis lattice vectors characterizing the torus $T^{2,j}$ and
${\alpha}^j$ is the angle between the
two basis vectors of the torus $T^{2,j}$. By performing a T-duality in the
$y$ direction of each torus $T^{2,j}$, the D9 brane with flux
$\mathcal{F}^j$ is converted to a D6 brane with an angle ${\theta}^j$ with
respect to the x-axis. This is given by \cite{Ardalan:1998ce}:
\begin{equation}
\tan(\pi\,{\theta}^j)=\frac{f^j}{\mathrm{Im}(T^{'j})}
\end{equation}
\noindent
where $f^j$ is defined by the quantization condition for the net 2-form
fluxes $\mathcal{F}^j$ as
\begin{equation}
f^j \equiv \frac{1}{(2\pi)^2{\alpha}'} \int_{T^{2,j}} \,
\mathcal{F}^j = \frac{m^j}{n^j}; \;\; m^j,n^j \,\in\,\mathbf{Z},
\end{equation}
\noindent
Using the above equation and the relation between the type $IIA$ and
type $IIB$:
\bea
T^j &=& -\frac{{\alpha}'}{U^{'j}},\;\;\;U^j =
-\frac{{\alpha}'}{T^{'j}},\;\;\;j=1,2,3. \\
R^j_1&=&R_1^{'j}, \qquad R^j_2=\frac{1}{R_2^{'j}}
\eea
\noindent we get the corresponding type $IIA$ relation:
\begin{equation}
\tan(\pi{\theta}^j) = \frac{m^j}{n^j}
\frac{|U^j|^2}{\mathrm{Im}\,(U^j)}; \;\; j=1,2,3. \label{idb:eq:a}
\end{equation}

\noindent The unprimed symbols correspond to the type IIA version
while the primed ones to the type IIB.

\subsection{$\mathcal{N}=1$ \textit{SUSY}}

We now look at the $\mathcal{N}=1$ supersymmetry constraint on the
open string sector.  In type $IIA$, this leads to a condition on
the angles ${\theta}^j$ \cite{Berkooz:1996npb}:
\begin{eqnarray}
{\theta}^1+{\theta}^2+{\theta}^3 &=& 0\;\; \mathrm{mod} \;\;2
\;\;\;\;
\mathrm{or} \nonumber \\
\sum_{j=1}^3 \frac{m^j}{n^j} \frac{|U^j|^2}{\mathrm{Im}\,(U^j)}
&=& \prod_{j=1}^3 \frac{m^j}{n^j}
\frac{|U^j|^2}{\mathrm{Im}\,(U^j)}
\end{eqnarray}

\noindent which after T-duality leads to a condition on the fluxes
in type $IIB$.

\section{Low energy effective action and soft terms}
\label{idb:set:soft}
We now analyze the issue of deriving the four dimensional
$\mathcal{N}=1$ low energy effective action of these intersecting
brane models. In the type $IIB$ picture, this has been done in
\cite{Grimm:2004uq,Jockers:2004yj} by Kaluza Klein
reduction of the Dirac-Born-Infeld and Chern-Simons action. The
effective action can also be obtained by explicitly computing disk
scattering amplitudes involving open string gauge and matter
fields as well as closed string moduli fields and deducing the
relevant parts of the effective action directly. This has been
done in \cite{Lust:2004cx}. We will follow the results of
\cite{Lust:2004cx} in our analysis.

The $\mathcal{N}=1$ supergravity action thus obtained is encoded
by three functions, the K\"{a}hler potential $K$, the
superpotential $W$ and the gauge kinetic function $f$
\cite{Cremmer:1982en}. Each of them will depend on the moduli
fields describing the background of the model. One point needs to
be emphasized. When we construct the effective action and its
dependence on the moduli fields, we need to do so in terms of the
moduli $s$, $t$ and $u$ in the field theory basis, in contrast to
the $S$, $T$ and $U$ moduli in the string theory basis
\cite{Lust:2004cx}. In type $IIA$, the real part of the field theory
$s$, $u$ and $t$ moduli are related to the corresponding string theory
$S$, $U$ and $T$ moduli by :
\begin{eqnarray}
\mathrm{Re}\,(s)& =&
\frac{e^{-{\phi}_4}}{2\pi}\,\left(\frac{\sqrt{\mathrm{Im}\,U^{1}\,
\mathrm{Im}\,U^{2}\,\mathrm{Im}\,U^3}}{|U^1U^2U^3|}\right)
\nonumber \\
\mathrm{Re}\,(u^j)& =&
\frac{e^{-{\phi}_4}}{2\pi}\left(\sqrt{\frac{\mathrm{Im}\,U^{j}}
{\mathrm{Im}\,U^{k}\,\mathrm{Im}\,U^l}}\right)\;
\left|\frac{U^k\,U^l}{U^j}\right| \qquad (j,k,l)=(\overline{1,2,3})
\nonumber \\
\re(t^j)&=&\frac{i\alpha'}{T^j} \label{idb:eq:moduli}
\end{eqnarray}

\noindent where $j$ stands for the $j^{th}$ 2-torus. The above
formulas can be inverted to yield the string theory $U$ moduli in
terms of the field theory moduli $s$ and $u$.
\begin{equation}
\frac{|U^j|^2}{\mathrm{Im}\,(U^j)} = \sqrt{
\frac{\mathrm{Re}\,(u^k)\,\mathrm{Re}\,(u^l)}{\mathrm{Re}\,(u^j)\mathrm{Re}\,
(s)}}\qquad  (j,k,l)=(\overline{1,2,3})
\label{eq:b}
\end{equation}

\noindent The holomorphic gauge kinetic function for a D6 brane
stack $P$ is given by :
\begin{eqnarray}
f_P &=& \frac{1}{\kappa_P}
(n_P^1\,n_P^2\,n_P^3\,s-n_P^1\,m_P^2\,m_P^3\,u^1-n_P^2\,m_P^1\,m_P^3\,u^2-
n_P^3\,m_P^1\,m_P^2\,u^3)
\nonumber \\
g_{D6_P}^{-2} &=& |\mathrm{Re}\,(f_P)|\label{idb:eq:gkf}
\end{eqnarray}
\noindent The extra factor ${\kappa}_P$ is related to the
difference between the gauge couplings for $U(N_P)$ and
$Sp(2N_P),\,SO(2N_P)$. ${\kappa}_P =1$ for $U(N_P)$ and
${\kappa}_P =2$ for $Sp(2N_P)$ or $SO(2N_P)$
\cite{Klebanov:2003my}.

\noindent The SM hypercharge $U(1)_Y$ gauge group is a linear
combination of several $U(1)$s: \be
Q_Y=\frac{1}{6}Q_a-\frac{1}{2}Q_c-\frac{1}{2}Q_d. \ee Therefore
the gauge kinetic function for the $U(1)_Y$ gauge group is
determined to be\cite{Blumenhagen:2003jy}: \be
f_Y=\frac{1}{6}f_{D_a}+\frac{1}{2}f_{D_c}+\frac{1}{2}f_{D_d}. \ee
The K\"{a}hler potential to the second order in open string matter
fields is given by :
\begin{eqnarray}
K(M,\bar{M},C,\bar{C}) = \hat{K}(M,\bar{M}) +  \sum_{untwisted\,
i,j} \tilde{K}_{C_{i}\bar{C}_j}(M,\bar{M}) C_i\bar{C}_j +
\sum_{twisted, \, \theta}
\tilde{K}_{C_{\theta}\bar{C}_{\theta}}(M,\bar{M})
C_{\theta}\bar{C}_{\theta}
\end{eqnarray}

\noindent where $M$ collectively denote the moduli; $C_i$ denote
untwisted open string moduli which comprise the D-brane positions
and the Wilson line moduli which arise from strings with both ends
on the same stack; and $C_{\theta}$ denote twisted open string
states arising from strings stretching between different stacks.

The open string moduli fields could be thought of as matter fields
from the low energy field theory point of view. The untwisted open
string moduli represent non-chiral matter fields and so do not
correspond to the MSSM. For the model to be realistic, they have
to acquire large masses by some additional mechanism, as already
explained in section \ref{idb:set:model:exo}.

Let's now write the K\"{a}hler metric for the twisted moduli
arising from strings stretching between stacks $P$ and $Q$, and
comprising $1/4$ BPS brane configurations. In the type IIA
picture, this is given by \cite{Font:2004cx,Cvetic:2003ch,Lust:2004cx}:
\begin{equation}
\tilde{K}_{C_{\theta_{PQ}}\bar{C}_{\theta_{PQ}}} \equiv
\tilde{K}_{PQ}= e^{{\phi}_4}\,\left(e^{{\gamma}_E\,\sum_{j=1}^{3}
{\theta}^j_{PQ}} \,\prod_{j=1}^3 \; \left[
\sqrt{\frac{{\Gamma}(1-{\theta}^j_{PQ})}{{\Gamma}({\theta}^j_{PQ})}}\;\;
(t^j+\bar{t}^j)^{-{\theta}^j_{PQ}}\right]\right)\label{idb:eq:kahler}
\end{equation}
\noindent where ${\theta}^j_{PQ} = {\theta}^j_{P}-{\theta}^j_{Q}$
is the angle between branes in the $j^{th}$ torus and ${\phi}_4$
is the four dimensional dilaton. From (\ref{idb:eq:moduli}),
${\phi}_4$ can be written as
$(\mathrm{Re}(s)\,\mathrm{Re}(u_1)\,\mathrm{Re}(u_2)\,\mathrm{Re}(u_3))^{-1/4}$.
The above K\"{a}hler metric depends on the field theory dilaton
and complex structure moduli $u^j$ through ${\phi}_4$ and
${\theta}_{PQ}$. It is to be noted that (\ref{idb:eq:kahler}) is a
product of two factors, one which explicitly depends on the field
theory $s$ and $u$ moduli ($e^{{\phi}_4}$), and the other which
implicitly depends on the $s$ and $u$ moduli (through the
dependence on ${\theta}^j_{PQ}$). Thus, $\tilde{K}_{PQ}$ can be
symbolically written as :\begin{equation} \tilde{K}_{PQ} =
e^{{\phi}_4}\,\tilde{K}^0_{PQ} \label{idb:eq:metric}
\end{equation}

\noindent The K\"{a}hler metric for $1/2$ BPS brane configurations
is qualitatively different from that of the $1/4$ BPS brane
configurations mentioned above. Generically, these cases arise if
both branes $P$ and $Q$ have a relative angle
${\theta}^j_{PQ}=0,1$ in the same complex plane $j$. It is
worthwhile to note that the higgs fields in Table
\ref{idb:tab:spec} form a $1/2$ BPS configuration and are
characterized by the following K\"{a}hler metric
\cite{Lust:2004fi}:
\begin{equation} \tilde{K}^{higgs}_{PQ} =
\left((s+\bar{s})(u^1+\bar{u}^1)(t^2+\bar{t}^2)(t^3+\bar{t}^3)\right)^{-1/2}
\label{idb:eq:higgs}
\end{equation}

\noindent An important point about the above expressions needs to
be kept in mind. These expressions for the K\"{a}hler metric has
been derived using \emph{tree level} scattering amplitudes and
with Wilson line moduli turned \emph{off}. Carefully taking the
Wilson lines into account as in \cite{Cremades:2004wa}, we see
that the K\"{a}hler metric has another multiplicative factor which
depends on the Wilson line moduli as well as $t$ moduli in type
$IIA$. If the Wilson line moduli do not get a vev, then our
analysis goes through. However, if they do, they change the
dependence of the metric on the $t$ moduli. As will be explained
later, we only choose the $u$ moduli dominated case for our
phenomenological analysis, so none of our results will be
modified.

The superpotential is given by:
\begin{equation}
W = \hat{W} + \frac{1}{2} {\mu}_{\alpha \beta}(M)\,
C^{\alpha}\,C^{\beta}+ \frac{1}{6}\,Y_{\alpha \beta
\gamma}(M)\,C^{\alpha}\,C^{\beta}\,C^{\gamma}+... \label{eq:W}
\end{equation}
In our phenomenological analysis, we have not included the Yukawa
couplings for simplicity. But as explained later, in the $u$
moduli dominant SUSY breaking case, the soft terms are independent
of the Yukawa couplings and will not change the phenomenology.

\subsection{Soft terms in general soft broken $\mathcal{N}=1$, $D=4$
supergravity Lagrangian}

From the gauge kinetic function, K\"{a}hler potential and the
superpotential, it is possible to find formulas for the
\emph{normalized} soft parameters - the gaugino mass parameters,
mass squared parameter for scalars and the trilinear parameters
respectively. These are given by \cite{Brignole:1997dp}:
\begin{eqnarray}
M_P &=& \frac{1}{2\,\mathrm{Re}\,f_P}\, (F^M\,\partial_M\,f_P) \nonumber \\
m_{PQ}^2 &=& (m_{3/2}^2 + V_0) - \sum_{M,N}\, \bar{F}^{\bar{M}}
F^N\,{\partial}_{\bar{M}}\,{\partial}_{N}\,\log({\tilde{K}}_{PQ}) \nonumber \\
A_{PQR} &=&
F^M[\hat{K}_M+{\partial}_M\,\log(Y_{PQR})-{\partial}_{M}\,
\log(\tilde{K}_{PQ}\tilde{K}_{QR}\tilde{K}_{RP})]\label{idb:eq:soft}
\end{eqnarray}

\noindent For our purposes, $P, Q$ and $R$ denote brane stacks. So
$M_P$ denotes the gaugino mass parameter arising from stack $P$;
$m_{PQ}^2$ denotes mass squared parameters arising from strings
stretching between stacks $P$ and $Q$ and $Y_{PQR}$ denotes Yukawa
terms arising from the triple intersection of stacks $P$, $Q$ and
$R$. The terms on the RHS without the indices $P$, $Q$ and $R$ are flavor
independent. Also, $M$ and $N$ run over the closed string moduli. $F^M$ stands
for the auxiliary fields of the moduli in general. Supersymmetry
is spontaneously broken if these fields get non-vanishing vevs. It
is assumed here that the auxiliary fields $D$ have vanishing vevs.
Their effect on the soft terms can be calculated as in
\cite{Kawamura:1996ex}, which we assume to be zero. These formulas
have been derived for the case when the K\"{a}hler metric for the
observable (MSSM) fields is diagonal and universal in flavor space. In principle,
there are also off-diagonal terms in the K\"{a}hler metric. They
relate twisted open string states at different intersections and
therefore are highly suppressed. We neglect the off- diagonal terms
in our study. If the
seperations between the intersections are very small, the off-diagonal
terms or non-universal diagonal terms
may have observable effects leading to interesting flavor
physics.

The effective $\mathcal{N} = 1$, $4\,d$ field theory is assumed to
be valid at some high scale, presumably the string scale. The
string scale in our analysis is taken to be the unification scale
$( \sim 2 \times 10^{16} \; \mathrm{GeV})$. We then need to use
the renormalization group equations (RGE) to evaluate these
parameters at the electroweak scale. In this paper, as mentioned
before, it is assumed that the non-chiral exotics have been made
heavy by some mechanism and there are no extra matter fields at
any scale between the electroweak scale and the unification scale.
This is also suggested by gauge coupling unification at the
unification scale.

One might wonder whether including the Yukawas in the analysis may
lead to significant modifications in the spectrum at low energies
because of their presence in the formulas for the soft terms
(\ref{idb:eq:soft}). However, this does not happen. This is
because the Yukawa couplings ($Y_{\alpha\beta\gamma}$) appearing in
the soft terms are \emph{not} the physical Yukawa couplings
($Y^{\mathrm{ph}}_{\alpha\beta\gamma}$). The two are related by:
\be
Y^{\mathrm{ph}}_{\alpha\beta\gamma} = Y_{\alpha\beta\gamma}\,
\frac{\hat{W}^*}{|\hat{W}|}\,e^{\hat{K}/2}\,
(\tilde{K}_{\alpha}\tilde{K}_{\beta}\tilde{K}_{\gamma})^{-1/2}
\ee
The Yukawa couplings
($Y_{\alpha\beta\gamma}$) between fields living at brane
intersections in intersecting D-brane models arise from worldsheet
instantons involving three different boundary conditions
\cite{Aldazabal:2000cn}. These semi-classical instanton amplitudes
are proportional to $e^{-A}$ where $A$ is the worldsheet area.
They have been shown to depend on the K\"{a}hler ($t$) moduli
(complexified area) and the Wilson line moduli
\cite{Cremades:2003qj} in type $IIA$. Although the physical Yukawas
($Y^{\mathrm{ph}}_{\alpha\beta\gamma}$) depend on the $u$ moduli through
their dependence on the K\"{a}hler potential, the fact that
$Y_{\alpha\beta\gamma}$ do not depend on the $u$ moduli in type $IIA$ ensures that in the
$F^u$ dominant supersymmetry breaking case, the soft terms are
independent of  $Y_{\alpha\beta\gamma}$.

Thus our analysis is similar in spirit to those in the case of the
heterotic string, where dilaton dominated supersymmetry breaking
and moduli dominated supersymmetry breaking are analyzed as
extreme cases. It should be remembered however, that T-duality
interchanges the field theory ($t$) and ($u$) moduli. Thus what we
call $u$ moduli in type $IIA$, become $t$ moduli in type $IIB$ and
vice versa. In a general situation, in which the $F$-terms of all
the moduli get vevs, the situation is much more complicated and a
more general analysis needs to be done. This is left for the
future.

\subsection{Soft terms in intersecting brane models}
We calculate the soft terms in type IIA picture. As already
explained, we assume that only $F$-terms for $u$ moduli get
non-vanishing vevs. In terms of the goldstino angle $\theta$ as
defined in \cite{Brignole:1997dp}, we have $\theta = 0$. We assume
the cosmological constant is zero and introduce the following
parameterization of $F^{u^i}$: \be
F^{u^i}=\sqrt{3}m_{3/2}(u^i+\bar u^i)\Theta_{i}e^{-i\gamma_i}
\qquad \mbox{for } i=1,2,3 \label{idb:eq:Fu} \ee with
$\sum|\Theta_i|^2=1$. To calculate the soft terms, we need to know
the derivatives of the K\"{a}hler potential with respect to $u$.
For a chiral field $C$ arising from open strings stretched between
stacks $P$ and $Q$, we denote its K\"{a}hler potential as
$\tilde{K}_{PQ}$. From (\ref{idb:eq:metric}), we see that their
derivatives with respect to $u^i$ are \bea \frac{\partial
\log{\tilde{K}_{PQ}}}{\partial u^i}&=& \sum_{j=1}^3\frac {\partial
\log{\tilde{K}^0_{PQ}}}{\partial\theta^j_{PQ}}
\frac{\partial\theta^j_{PQ}}{\partial u^i} + \frac{-1}{4(u^i+\bar{u}^i)}\\
\frac{\partial^2 \log{\tilde{K}_{PQ}}}{\partial u^i\partial\bar
u^j}&=& \sum_{k=1}^3\left(\frac{\partial
\log{\tilde{K}^0_{PQ}}}{\partial\theta^k_{PQ}}
\frac{\partial^2\theta^k_{PQ}}{\partial u^i\partial\bar u^j}+
\frac{\partial^2\log \tilde{K}^0_{PQ}}{\partial (\theta^k_{PQ})^2}
\frac{\partial\theta^k_{PQ}}{\partial u^i}
\frac{\partial\theta^k_{PQ}}{\partial \bar
u^j}+\frac{{\delta}_{ij}}{4\,(u^i+\bar{u}^i)^2}\right) \eea From
the K\"{a}hler potential in eq.(\ref{idb:eq:kahler}), we have \bea
\Psi(\theta^j_{PQ})&\equiv&\frac {\partial
\log{\tilde{K}^0_{PQ}}}{\partial\theta^j_{PQ}}=
\gamma_E+\frac{1}{2}\frac{d}{d{\theta}^j_{PQ}}\,\ln{\Gamma(1-\theta^j_{PQ})}-
\frac{1}{2}\frac{d}{d{\theta}^j_{PQ}}\,\ln{\Gamma(\theta^j_{PQ})}-\log(t^j+\bar t^j) \label{eq:idb:gamma}\\
\Psi'(\theta^j_{PQ})&\equiv& \frac{\partial^2\log
\tilde{K}^0_{PQ}} {\partial(\theta^j_{PQ})^2}=
\frac{d\Psi(\theta^j_{PQ})}{d \theta^j_{PQ}} \eea The angle
$\theta^j_{PQ}\equiv\theta^j_P-\theta^j_Q$ can be written in terms
of $u$ moduli as: \be \tan(\pi\theta^j_P)=\frac{m^j_P}{n^j_P}
\sqrt{\frac{\re u^k \re u^l}{\re u^j\re s}} \qquad\mbox{where }
(j,k,l)=(\overline{1,2,3}) \ee Then we have \be
{\theta}^{j,k}_{PQ} \equiv (u^k+\bar u^k)\,\frac{\partial
\theta^j_{PQ}}{\partial u^k}= \left\{\begin{array}{cc}
 \left[-\frac{1}{4\pi}
 \sin(2\pi\theta^j)
 \right]^P_Q & \mbox{ when }j=k  \vspace*{0.6cm} \\
 \left[\frac{1}{4\pi}
\sin(2\pi\theta^j)
 \right]^P_Q & \mbox{ when }j\neq k
\end{array}\right.\label{idb:eq:dthdu}
\ee where we have defined
$[f(\theta^j)]^P_Q=f(\theta^j_P)-f(\theta^j_Q)$. The second order
derivatives are \be {\theta}^{j,k\bar{l}}_{PQ} \equiv (u^k+\bar
u^k)(u^l+\bar u^l)\,\frac{\partial^2 \theta^j_{PQ}}{\partial
u^k\partial\bar u^l}= \left\{\begin{array}{cc} \frac{1}{16\pi}
  \left[ \sin(4\pi\theta^j)+4\sin(2\pi\theta^j)
 \right]^P_Q &
   \mbox{when }j=k=l  \vspace*{0.6cm} \\
 \frac{1}{16\pi}  \left[
 \sin(4\pi\theta^j)-4\sin(2\pi\theta^j)
 \right]^P_Q &
   \mbox{when }j\neq k=l  \vspace*{0.6cm} \\
 -\frac{1}{16\pi}\left[
 \sin(4\pi\theta^j)
 \right]^P_Q &
   \mbox{ when }j=k\neq l\mbox{ or } j=l\neq k \vspace*{0.4cm} \\
 \frac{1}{16\pi}\left[
\sin(4\pi\theta^j)
 \right]^P_Q &
   \mbox{when }j\neq k\neq l\neq j
\end{array}\right.\label{idb:eq:dth2du}
\ee Now, one can substitute
eq.(\ref{idb:eq:Fu}-\ref{idb:eq:dth2du}) to the general formulas
of soft terms in eq.(\ref{idb:eq:soft}). The formulas for the soft
parameters in terms of the wrapping numbers of a general
intersecting D-brane model are given by:

\begin{itemize}
\item Gaugino mass parameters \be M_P=\frac{-\sqrt{3}m_{3/2}}{\re
f_P}\sum_{j=1}^3 \left(\re u^j\,\Theta_j\,
e^{-i\gamma_j}\,n^j_Pm^k_Pm^l_P\right) \qquad
(j,k,l)=(\overline{1,2,3}) \label{eq:idb:gaugino}\ee
As an example, the $M_{\tilde{g}}$ can be obtained by putting $P = a$ and
using the appropriate wrapping numbers, as in
Table \ref{idb:tab:wr}.

\item Trilinears : \begin{eqnarray} A_{PQR}&=&-\sqrt{3}m_{3/2}\sum_{j=1}^3 \left[
\Theta_je^{-i\gamma_j}\left(1+(\sum_{k=1}^3
 \theta_{PQ}^{k,j}\Psi(\theta^k_{PQ})-\frac{1}{4})+(\sum_{k=1}^3
 \theta_{RP}^{k,j}\Psi(\theta^k_{RP})-\frac{1}{4})
\right)\right] + \nonumber \\
& & \frac{\sqrt{3}}{2}m_{3/2}{\Theta}_{1}e^{-i{\gamma}_1}\end{eqnarray}
This arises in general from the triple intersections $PQ$, $QR$
and $RP$, where $PQ$ and $RP$ are 1/4 BPS sector states and $QR$ is a 
1/2 BPS state. $QR$ being the higgs field, has a special contribution
to the trilinear term compared to $PQ$ and $RP$. So as an example,
$A_{Q_LH_uU_R}$ can be obtained as a triple intersection of $ab$,
$bc$ and $ca$, as seen from Table \ref{idb:tab:spec}.

\item Scalar mass squared (1/4 BPS) : \be
m^2_{PQ}=m_{3/2}^2\left[1-3\sum_{m,n=1}^3
\Theta_m\Theta_ne^{-i(\gamma_m-\gamma_n)}\left(
\frac{{\delta}_{mn}}{4}+ \sum_{j=1}^3 (\theta^{j,m\bar
n}_{PQ}\Psi(\theta^j_{PQ})+
 \theta^{j,m}_{PQ}\theta^{j,\bar n}_{PQ}\Psi'(\theta^j_{PQ}))\right)
\right] \ee
By using the definitions in eq.(\ref{idb:eq:dthdu},
\ref{idb:eq:dth2du}), we see that $\theta^{j,k}_{PQ}$ and
$\theta^{j,k\bar l}_{PQ}$ do not depend on  the vevs of $u$
moduli: $u^j$. As an example, the squark mass squared
$m^2_{\tilde{Q}}$ can be obtained by putting $P,Q = a,b$; as can
be seen again from Table \ref{idb:tab:spec}.
\end{itemize}

\noindent We can now use the wrapping numbers in Table
\ref{idb:tab:wr} to get explicit expressions for the soft terms
for the particular model in terms of the $s,t$ and $u$ moduli and
the parameters $m_{3/2}$, $\gamma_i$ and $\Theta_i$, $i=1,2,3$.
The expressions for the trilinear $A$ parameters and scalar mass
squared parameters (except those for the up and down type higgs)
are provided in the Appendix. Using Table \ref{idb:tab:wr}, the
formula for the gaugino mass parameters and the mass squared
parameters for the up and down higgs are given by:

\begin{itemize}
\item Gaugino mass parameters: \bea
M_{\wtd B}&=&3\sqrt{3}m_{3/2}
 \frac{12e^{-i\gamma_1}\,\re u_1\,\Theta_1
 +e^{-i\gamma_3}\,\re u_3\,\Theta_3}
 {3\re u_3+4\re s+ 36 \re u_1} \\
M_{\wtd W}&=&\sqrt{3}m_{3/2}\,e^{-i\gamma_2}\,\Theta_2 \\
M_{\wtd g}&=&\sqrt{3}m_{3/2}\frac{9e^{-i\gamma_1}\,\re
u_1\,\Theta_1}
 {\re s+9\re u_1}
\eea
It is important to note that there is no gaugino mass degeneracy at the
unification scale, unlike other models such as mSUGRA. This will lead to
interesting experimental signatures.

\item Higgs mass parameters \bea m^2_{H_u} = m^2_{H_d} =
m^2_{3/2}\,(1-\frac{3}{2}\,|{\Theta}_1|^2)\eea

\noindent The Higgs mass parameters are equal because they are
characterized by the same K\"{a}hler metric
$\tilde{K}^{higgs}_{PQ}$ in (\ref{idb:eq:higgs}).

\end{itemize}

\noindent For completeness, we would also like to list the
relative angles between a brane $P$ and the orientifold plane on
the $j$th torus are denoted by $\theta^j_P$ in Table \ref{idb:tab:angl}. 
The soft terms depend on these angles.

\begin{table}
\begin{center}
\begin{tabular}[h]{|c|c|c|c|c|c|}
\hline Stack  & Number of Branes & Gauge Group & $({\theta}^1_P)$
&
$({\theta}^2_P)$ & $({\theta}^3_P)$\\
\hline $Baryonic$ & $N_a =3$ & $U(3)= SU(3) \times U(1)_a$ &
$0$ & $\alpha$ & $-\alpha$ \\
$Left$ & $N_b =1$ & $USp(2)\cong SU(2)$ & $\frac{1}{2}$
 & $0$ & $-\frac{1}{2}$\\
$Right$ & $N_c =1$ & $U(1)_c$ & $\frac{1}{2}$ &
 $-\frac{1}{2}$ & $0$\\
$Leptonic$ & $N_d =1$ &$U(1)_d$ &$0$ & $\alpha$ & $-\alpha$
\\
\hline
\end{tabular}
\end{center}
\caption{Angles between a brane $P$ and the orientifold plane on
the $j$th torus: $\theta^j_P$.} \label{idb:tab:angl}
\end{table}

\vspace{0.3cm}

\noindent The only non-trivial angle is $\alpha$, which is given
by \be \tan (\pi\alpha) =3\,\rho^2\sqrt{\frac{\re u^1\re u^3}{\re
u^2 \re s}}. \ee as can be seen from
eq.(\ref{idb:eq:a},\ref{eq:b}) and Table \ref{idb:tab:wr}.

We would now like to compare our setup and results for the soft
terms with those obtained in \cite{Lust:2004dn}. The setup
considered in \cite{Lust:2004dn} is very similar to ours, but with
a few differences. It is a type $IIB$ setup with three stacks of
D7 branes wrapping four cycles on $T^6$. The last stack is
equipped with a non-trivial open string flux, to mimic the intersecting
brane picture in type $IIA$, as explained in section 3.4. There is
also a particular mechanism of supersymmetry breaking through
closed string 3-form fluxes. Thus, there is an explicit
superpotential generated for the closed string moduli, which leads
to an explicit dependence of the gravitino mass $(m_{3/2})$,
$F^{s,t^i,u^i}$ and the cosmological constant $(V)$ on the moduli
$s$, $t^i$ and $u^i$. The cosmological constant is zero if the
goldstino angle ($\theta$) is zero which is the same in our case.
It turns out that using these formulas, in order for the gravitino
mass to be small, the string scale is sufficiently
low for reasonable values
($\mathcal{O}(1)$ in string units) of the flux. We have not
assumed any particular mechanism of supersymmetry breaking, so we do
not have an explicit expression for $(m_{3/2})$, $F^{s,t^i,u^i}$
and $(V)$ in terms of the moduli and have taken the string scale
to be of the order of the unification scale.

The model considered in \cite{Lust:2004dn}
considers non-zero (0,3) and (3,0) form fluxes only, which leads
to non-vanishing $F^{s}$ and $F^{t^i}$. In the T-dual version,
this means that $F^{s}$ and $F^{u^i}$ are non-zero, which is the
case we examined in detail. However, in \cite{Lust:2004dn}, an
isotropic compactification is considered, while we allow a more general situation.

For the calculation of soft terms, we have used the updated form of the 
1/4 BPS sector K\"{a}hler metric as in \cite{Font:2004cx}, which we have also 
explicitly checked. In \cite{Lust:2004dn}, the \emph{un-normalized} general expression
for calculating the soft terms has been used following
\cite{Kaplunovsky:1993rd}, whereas we use the \emph{normalized}
general expression for the soft terms in eq.(\ref{idb:eq:soft}) \cite{Brignole:1997dp}.  
In contrast to \cite{Lust:2004dn} which has a left-right symmetry, 
we have also provided an expression for the Bino mass parameter, since we have the SM 
gauge group (possibly augmented by $U(1)'s$) and the exact linear combination of $U(1)'s$ giving rise 
to $U(1)_Y$ is known.

\section{Some phenomenological implications}
\label{idb:set:phen}
Using the formulas for the soft terms given in the previous
section, we can study some aspects of the phenomenology of the
model that has the brane setup shown in Table \ref{idb:tab:wr}.

Although ideally the theory should generate $\mu$ of order the
soft terms and $\tan\beta$ should be calculated, that is not yet
possible in practice as explained before. Therefore we will not
specify the mechanism for generating the $\mu$ term. We will take
$\tan\beta$ and $m_Z$ as input and use EWSB to fix $\mu$ and $b$
terms.

Unlike the heterotic string models where the gauge couplings are
automatically unified\footnote{But usually unified at a higher
scale than the true GUT scale.}, generic brane models don't have
this nice feature. This is simply because in brane models
different gauge groups originate from different branes and they do
not necessarily have the same volume in the compactified space.
Therefore to ensure gauge coupling unification at the scale
$M_X\approx 2 \times 10^{16}$GeV, the vev of some moduli fields
need to take certain values. In our models, the gauge couplings
are determined according to eq.(\ref{idb:eq:gkf}). Thus the
unification relations
\be g_s^2=g_w^2=\frac{5}{3}g_Y^2\approx 0.5.
\label{idb:eq:guni}
\ee
lead to  three conditions on the four
variables: $\re\, s$ and $\re\, u^i$ where $i=1,2,3$. One of the
solutions is
\be \re s=2-9\re u_1 \qquad \re u_2=4 \qquad
\re u_3=4.
\ee
It's interesting to note that $\mathcal{N}=1$ SUSY
condition actually requires $\re\,u_2=\re\, u_3$.
Therefore although at first sight it seems that the three gauge
couplings are totally unrelated in brane models, in this case
requiring $\mathcal{N}=1$ SUSY actually guarantees one of the
gauge coupling unification conditions \cite{Blumenhagen:2003jy}.

After taking into account the gauge coupling unification
constraint, the undetermined parameters we are left with are
$m_{3/2}$, $\tan\beta$, $\re u^1$, $\re t^2$, $\re t^3$, $\Theta_i$
and $\gamma_i$, where $i=1,2,3$.
The phases $\gamma_{1,2,3}$ enter both gaugino masses and trilinears and
in general can not be rotated away, leading to EDMs and a
variety of other interesting phenomena for colliders, dark matter,
etc. in principle. However, for simplicity, in this paper we set all
phases to be zero: $\gamma_{1,2,3}=0$.
The only dependence on
$\re\, t^2$ and $\re\, t^3$ is in the scalar mass squared terms
and is logarithmic. For simplicity, we set them equal: $\re t^2=\re t^3$.
Using the relation $\sum|\Theta_i|^2=1$, we can eliminate the magnitude
of one of the $\Theta_i$s but its sign is free to vary.
Thus  we are left with the following six free parameters and two signs:
\be m_{3/2},\hspace*{0.3cm}\tan\beta,
\hspace*{0.3cm}\re u^1, \hspace*{0.3cm}\re t^2,
\hspace*{0.3cm}\Theta_1, \hspace*{0.3cm}\Theta_2,
\hspace*{0.3cm}\mbox{sign}(\mu), \hspace*{0.3cm} \mbox{sign}(\Theta_3).
\ee
Instead of
scanning the full parameter space, we show here three representative
models which correspond to three interesting points in the parameter space. In
order to reduce fine tuning, we require that the gluino be not too
heavy. We also require that the higgs mass is about 114 GeV, that all
other experimental bounds are satisfied and that the universe is not
overclosed. These requirements strongly constrains
the free parameters. The parameters of these three points
are shown in Table \ref{idb:tab:para}.
\begin{table}[h]
\begin{center}
\begin{tabular}{|c|c|c|c|c|c|c|c|c|c|} \hline
LSP &$m_{3/2}$ & $\tan\beta$&$\re u^1$&$\re t^2$&$\Theta_1$& $\Theta_2$
 & $\mbox{sign}(\mu)$ & $\mbox{sign}(\Theta_3)$ &
  $\Omega^{\mbox{thermal}}_{\wtd N_1}$ \\ \hline
$\wtd W$  & 1500& 20& 0.025& 0.01& 0.50& 0.060& $+$& $-$& $\sim
0$\\ \hline $\wtd H$  & 2300& 30& 0.025& 0.01& -0.75& 0.518& $+$&
$+$& $\sim 0$\\ \hline $\wtd B$-$\wtd H$ mixture  &
               2300& 30& 0.025& 0.01& -0.75& 0.512& $+$& $+$& $\sim 0.23$ \\ \hline
\end{tabular}
\end{center}
\caption{Parameter choices for three particular models. All masses
are in GeV. We set $\re t^3=\re t^2$. $|\Theta_3|$ will be fixed by
the condition $\sum|\Theta_i|^2=1$. $\re s$, $\re u^2$ and $\re u^3$
are determined by
requiring gauge coupling unification. } \label{idb:tab:para}
\end{table}

Using the values of the moduli, one can calculate the string scale
$M_S$. It is indeed between the unification scale and the Planck scale.
Notice that since $t\sim 1/T$, we are in the large radius limit
of compactification and perturbation theory holds good.

From the parameters shown in Table \ref{idb:tab:para}, we
can calculate the soft terms at high scale. They
are shown in Table \ref{idb:tab:hsoft}.
\begin{table}[h]
\begin{center}
\begin{tabular}{|c|c|c|c|c|c|} \hline
 & $M_1$ & $M_2$ & $M_3$ &  $A_t$& $m_0$ \\ \hline
$\wtd W$ LSP &-1288& 156& 146 & -728 & $\sim m_{3/2}$  \\ \hline
$\wtd H$ LSP &849&2064& -336 & 633 & $\sim m_{3/2}$   \\ \hline
$\wtd B$-$\wtd H$ &
               866& 2040& -336& 640 & $\sim m_{3/2}$   \\ \hline
\end{tabular}
\end{center}
\caption{Soft terms at the unification scale. The input parameters
for calculating the soft terms are shown in Table
{\ref{idb:tab:para}}. $m_0$ denotes the average of scalar masses.
In both models, they are roughly the gravitino mass. The sign of the trilinears
is according to the convention used by SUSPECT. It should be kept in mind that this is 
opposite to the convention used in supergravity formulas.}
\label{idb:tab:hsoft}
\end{table}
We use SUSPECT \cite{Djouadi:2002ze} to run the soft terms from
the high scale to the weak scale and calculate the sparticle
spectrum. Most string-based models that have been analyzed in such
detail have had Bino LSPs.
The three models we examine give wino, higgsino and mixed bino-higgsino
LSP, respectively. The gluino masses\footnote{The radiative
contributions to the gluino pole mass from squarks are included.}
in the three models are $\mathcal{O}(500\mbox{GeV})$.
They are significantly lighter than the gluinos in
most of existing supergravity and superstring based MSSM models.
These three models satisfy all current experimental constraints.
If the LSP is a pure bino, usually the relic density is too large
to satisfy the WMAP\cite{Eguchi:2002dm}
 constraint. For the third model, the LSP is a
mixture of bino and higgsino such that its thermal relic density
is just the measured $\Omega_{CDM}$.
For the first two models,  LSPs annihilate quite efficiently such that
their thermal relic density is negligible.
Thus if only thermal production is taken into account,
the LSP can not be the cold dark matter (CDM) candidate. But in
general, there are non-thermal production mechanisms for the LSP,
for example the late gravitino decay\cite{Kawasaki:1994af}
or Q-ball decays \cite{Fujii:2001xp,Kasuya:1999wu}. These non-thermal production
mechanisms have uaually been neglected but could have
important effects on predicting the relic density of the LSP. Since at
present, it is not known whether non-thermal production mechansims
were relevant in the early universe, we have presented examples of both
possibilities.

Late gravitino decay actually can not generate enough LSPs to
explain the observed $\Omega_{CDM}$ for the first two models.
This is because
the gravitino decays after nucleosynthesis. Thus to
avoid destroying the successful nucleosynthesis predictions, the
gravitino should not be produced abundantly during the reheating
process. Therefore, LSPs from the gravitino decay can not be the
dominant part of CDM.

Another mechanism for non-thermal production is by the late Q-ball
decay. Q-balls are non-topological solitons\cite{Coleman:1985ki}.
At a classical level, their stability is guaranteed by some conserved
charge, which could be associated with either a global or a local
$U(1)$ symmetry. At a quantum level, since the field configuration
corresponding to Q-balls does not minimize the potential globally
and in addition Q-balls by definition do not carry conserved
topological numbers, they will ultimately decay. Q-balls can be
generated in the Affleck-Dine mechanism of
baryogenesis\cite{Dine:1995kz}. Large amounts of
supersymmetric scalar particles can be stored in Q-balls. Final
Q-ball decay products will include LSPs, assuming R-parity is
conserved. If Q-balls decay after the LSP freezes out and before
nucleosynthesis, the LSP could be the CDM candidate and explain
the observed $\Omega_{CDM}$. The relic density
of $\wtd N_1$ can be estimated as\cite{Fujii:2001xp}
\be
\Omega_{\wtd N_1}\approx 0.5\times\left(\frac{0.7}{h}\right)^2
\times\left(\frac{m_{\wtd N_1}}{100\mbox{GeV}}\right)
\left(\frac{10^{-7}\mbox{GeV}^2}{\langle\sigma v\rangle }\right)
\left(\frac{100\mbox{MeV}}{T_d}\right)
\left(\frac{10}{g_*(T_d)}\right)^{1/2}
\ee
For our first two models,
$\langle\sigma v\rangle\sim 10^{-7}\mbox{GeV}^2$. Thus if the
temperature $(T_d)$ when Q-balls decay is about $100$
MeV, we will have $\Omega_{\wtd N_1}\approx 0.22$. One may attempt to
relate this number to the baryon number of the universe since
Q-balls may also carry baryon number. But probably, this wouldn't
happen in these models at least at the perturbative level, because
then baryon number is a global symmetry.
Therefore Q-balls wouldn't carry net baryon number. They may carry
lepton number because of lepton number violation. But since $T_d$ is well
below the temperature of the electroweak (EW) phase transition,
sphaleron effects are well suppressed so that the net lepton
number cannot be transferred to baryon number. Hence, in the
Q-ball scenario, baryon number probably is not directly related to
$\Omega_{\wtd N_1}$.

\begin{figure}
\center \epsfig{file=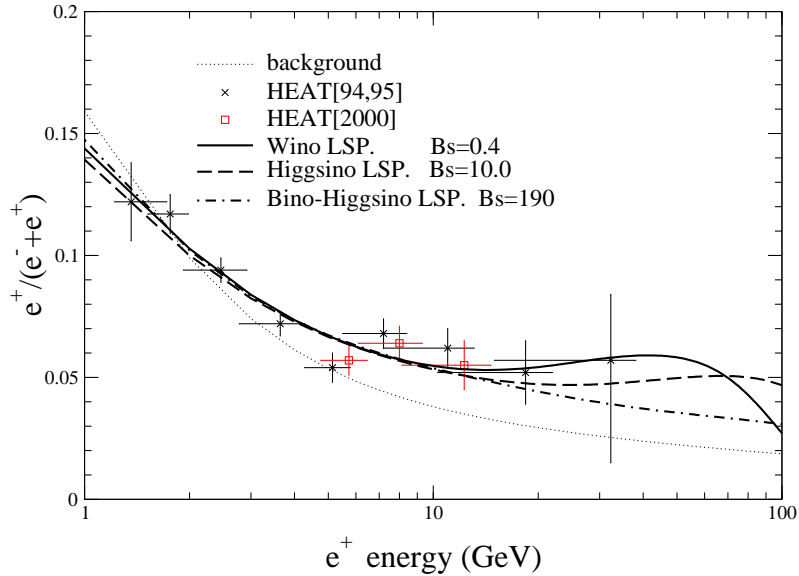,height=12cm, angle=-90}
\caption{Cosmic ray positron excess due to LSP annihilations.
Crossed points are HEAT experimental data. The dotted line is from the
standard cosmological model without taking into account the LSP
annihilation contributions. The solid line is for the $\wtd W$
LSP model, the dashed line is for the $\wtd H$ LSP model and the dashed
-dotted line is for the mixed $\wtd B$-$\wtd H$ LSP model. Bs is the
boost factor.} \label{idb:fig:posi}
\end{figure}

If we assume $\Omega_{\wtd N_1}\approx 0.23$ by either thermal (for the third
model) or non-thermal production (for the first two models),
then there will be interesting experimental signatures.
One of them is a positron excess in cosmic rays.
Such an excess has been reported by the HEAT experiment\cite{Barwick:1995gv}
and two other experiments (AMS and PAMELA)
\cite{Lechanoine-Leluc:2004dm} have been planned
which will give improved results in the future. In fact the HEAT excess
is currently the only reported dark matter signal that is
consistent with other data.
In both the wino LSP and the higgsino LSP models, LSPs annihilate to $W$
bosons quite efficiently. In particular, $W^+$ decays to a positron(and
a neutrino).
We used DARKSUSY\cite{Gondolo:2000ee} to fit the HEAT experiment
measurements\cite{Kane:2002nm}. The fitted curves are shown in
Figure \ref{idb:fig:posi}. The ``boost'' factors,
which characterize the local CDM density fluctuation against the
averaged halo CDM density,  in both models are
not large. Small changes in other parameters could give boost
factors very near unity. Thus
both models give a nice explanation for the
measured positron excess in the cosmic ray data.
For the mixed LSP model, because the LSPs do not annihilate efficiently,
one needs a ``boost'' factor of $\mathcal{O}(100)$ to get a good
fit, which may be a bit too high.

\section{Conclusion}\label{idb:sec:conc}
In this paper we investigated some phenomenological implications
of intersecting D-brane models, with emphasis on dark matter. We
calculated the soft SUSY breaking terms in these models focussing
on the $u$ moduli dominated SUSY breaking scenario in type $IIA$,
in which case the results do not depend on the Yukawa couplings
and Wilson lines. The results depend on the brane wrapping numbers
as well as SUSY breaking parameters. Our main result is providing
in detail the soft-breaking Lagrangian for intersecting brane models,
which provides
a new set of soft parameters to study phenomenologically.
We use a rather general parameterization of F-term vevs,
based on \cite{Brignole:1997dp}, which can be specialized to
include the recent progress on susy breaking by flux compactification
\cite{Camara:2003ku,Lust:2004fi,Lust:2004dn}.
Our results overlap and are consistent with those of
\cite{Lust:2004dn}. We applied our results to a particular
intersecting brane model\cite{Cremades:2003qj} which gives an MSSM-like particle
spectrum, and then selected three representative points in the
parameter space with relatively light gluinos, in order to reduce
fine-tuning,
and calculated the weak scale spectrum for them.
The phenomenology of the three models corresponding to the three
points is very interesting. The LSPs have different properties.
They can be either wino-like, higgsino-like, or a mixture of bino and higgsino.
All of them can be good candidates for the CDM.

\section{Appendix}

Here, we list the formulas for the soft scalar mass parameters and
the trilinear parameters. It turns out that for the matter content
in Table 2, we have the following independent soft parameters :

\begin{itemize}
\item \underline{Trilinear parameters ($A$)}:
\begin{eqnarray}
A_{Q_LH_uU_R} &=&
\frac{-\sqrt{3}m_{3/2}}{4\pi}\,[ 2\pi({\Theta}_2+{\Theta}_3)+[({\Theta}_3-{\Theta}_2) 
\left( \Psi({\theta}^2_{ab})+\Psi({\theta}^3_{ab})-\Psi({\theta}^2_{ca})-\Psi({\theta}^3_{ca}) \right) \nonumber \\
& & + {\Theta}_1 \, \left( \Psi({\theta}^2_{ab})-\Psi({\theta}^3_{ab})-\Psi({\theta}^2_{ca})+\Psi({\theta}^3_{ca}) \right)]
\,\sin(2\pi\alpha)\,] \\
A_{Q_LH_dD_R} &=&
-\frac{\sqrt{3}m_{3/2}}{4\pi}\,[ 2\pi({\Theta}_2+{\Theta}_3)+[({\Theta}_3-{\Theta}_2) 
\left( \Psi({\theta}^2_{ab})+\Psi({\theta}^3_{ab})-\Psi({\theta}^2_{c*a})-\Psi({\theta}^3_{c*a}) \right) \nonumber \\
& & + {\Theta}_1 \, \left( \Psi({\theta}^2_{ab})-\Psi({\theta}^3_{ab})-\Psi({\theta}^2_{c*a})+\Psi({\theta}^3_{c*a}) \right)]
\,\sin(2\pi\alpha)\,] \\
A_{Q_LH_dD_R} &=& A_{LH_dE_R};\;\; A_{Q_LH_uU_R} = A_{LH_uN_R} \nonumber 
\end{eqnarray}
\newpage
\item \underline{Scalar Mass parameters (${\tilde{m}}^2$)}:
\begin{eqnarray}
{\tilde{m}}^2_{Q_L} &=& \frac{-m_{3/2}^2}{16{\pi}^2}\;[-12\pi\,\{{{\Theta}_1}^2({\Psi}({\theta}^2_{ab})
-{\Psi}({\theta}^3_{ab})) - ({{\Theta}_2}^2-{{\Theta}_3}^2)
({\Psi}({\theta}^2_{ab})+{\Psi}({\theta}^3_{ab}))\}\,\sin(2\pi\alpha) + \nonumber \\
& &  3 \,\{-2 {\Theta}_1 ({\Theta}_2-{\Theta}_3) ({\Psi}'({\theta}^2_{ab})-{\Psi}'({\theta}^3_{ab})) +     
{{\Theta}_1}^2 ({\Psi}'({\theta}^2_{ab})+{\Psi}'({\theta}^3_{ab})) + \nonumber \\
& & ({\Theta}_2-{\Theta}_3)^2 ({\Psi}'({\theta}^2_{ab})+{\Psi}'({\theta}^3_{ab}))\}
\, {\sin}^2(2\pi\alpha) + \pi \,[(-4\pi) + \nonumber \\ 
& & 3 \,\{{\Theta}_1^2 ({\Psi}({\theta}^2_{ab})-{\Psi}({\theta}^3_{ab})) + 
({\Theta}_2-{\Theta}_3)^2 ({\Psi}({\theta}^2_{ab})-{\Psi}({\theta}^3_{ab})) 
- \nonumber \\
& & 2 {\Theta}_1 ({\Theta}_2-{\Theta}_3) ({\Psi}({\theta}^2_{ab})+{\Psi}({\theta}^3_{ab}))\}\,\sin(4\pi\alpha) \,]\;] \\
{\tilde{m}}^2_{U_R} &=& \frac{-m_{3/2}^2}{16{\pi}^2}\;[-12\pi\,\{{{\Theta}_1}^2({\Psi}({\theta}^2_{ac})
-{\Psi}({\theta}^3_{ac})) - ({{\Theta}_2}^2-{{\Theta}_3}^2)
({\Psi}({\theta}^2_{ac})+{\Psi}({\theta}^3_{ac}))\}\,\sin(2\pi\alpha) + \nonumber \\
& &  3 \,\{-2 {\Theta}_1 ({\Theta}_2-{\Theta}_3) ({\Psi}'({\theta}^2_{ac})-{\Psi}'({\theta}^3_{ac})) +     
{{\Theta}_1}^2 ({\Psi}'({\theta}^2_{ac})+{\Psi}'({\theta}^3_{ac})) + \nonumber \\
& & ({\Theta}_2-{\Theta}_3)^2 ({\Psi}'({\theta}^2_{ac})+{\Psi}'({\theta}^3_{ac}))\}
\, {\sin}^2(2\pi\alpha) + \pi \,[(-4\pi) + \nonumber \\ 
& & 3 \,\{{\Theta}_1^2 ({\Psi}({\theta}^2_{ac})-{\Psi}({\theta}^3_{ac})) + 
({\Theta}_2-{\Theta}_3)^2 ({\Psi}({\theta}^2_{ac})-{\Psi}({\theta}^3_{ac})) 
- \nonumber \\
& & 2 {\Theta}_1 ({\Theta}_2-{\Theta}_3) ({\Psi}({\theta}^2_{ac})+{\Psi}({\theta}^3_{ac}))\}\,\sin(4\pi\alpha) \,]\;] \\
{\tilde{m}}^2_{D_R} &=& \frac{-m_{3/2}^2}{16{\pi}^2}\;[-12\pi\,\{{{\Theta}_1}^2({\Psi}({\theta}^2_{ac*})
-{\Psi}({\theta}^3_{ac*})) - ({{\Theta}_2}^2-{{\Theta}_3}^2)
({\Psi}({\theta}^2_{ac*})+{\Psi}({\theta}^3_{ac*}))\}\,\sin(2\pi\alpha) + \nonumber \\
& &  3 \,\{-2 {\Theta}_1 ({\Theta}_2-{\Theta}_3) ({\Psi}'({\theta}^2_{ac*})-{\Psi}'({\theta}^3_{ac*})) +     
{{\Theta}_1}^2 ({\Psi}'({\theta}^2_{ac*})+{\Psi}'({\theta}^3_{ac*})) + \nonumber \\
& & ({\Theta}_2-{\Theta}_3)^2 ({\Psi}'({\theta}^2_{ac*})+{\Psi}'({\theta}^3_{ac*}))\}
\, {\sin}^2(2\pi\alpha) + \pi \,[(-4\pi) + \nonumber \\ 
& & 3 \,\{{\Theta}_1^2 ({\Psi}({\theta}^2_{ac*})-{\Psi}({\theta}^3_{ac*})) + 
({\Theta}_2-{\Theta}_3)^2 ({\Psi}({\theta}^2_{ac*})-{\Psi}({\theta}^3_{ac*})) 
- \nonumber \\
& & 2 {\Theta}_1 ({\Theta}_2-{\Theta}_3) ({\Psi}({\theta}^2_{ac*})+{\Psi}({\theta}^3_{ac*}))\}\,\sin(4\pi\alpha) \,]\;] \\
{\tilde{m}}^2_{Q_L} &=&  {\tilde{m}}^2_{L}; \;\; {\tilde{m}}^2_{U_R} = {\tilde{m}}^2_{N_R}; \;\; {\tilde{m}}^2_{D_R} =  {\tilde{m}}^2_{E_R} \nonumber
\end{eqnarray}
\end{itemize}

\noindent The above formulas are subject to the constraint $|{\Theta}_1|^2+|{\Theta}_2|^2+|{\Theta}_3|^2 = 1$, as can be 
seen below (\ref{idb:eq:Fu}).

\vspace*{1cm}

\noindent{\Large \bf Acknowledgement} \\
The authors appreciate helpful conversations with and
suggestions from Brent~D.~Nelson, \\ Andrew Pawl
and Lian-Tao~Wang. The research of GLK, PK, JDL and TTW are supported in 
part by the US Department of Energy.

\end{document}